\documentclass[vecphys]{svmult}

\usepackage{makeidx}         
\usepackage{graphicx}        
\usepackage{epsfig}           
\usepackage{multicol}        
\usepackage{url}             

\makeindex             

\def\be{\begin{equation}}
\def\ee{\end{equation}}
\def\lsim{\lower 2pt \hbox{$\, \buildrel {\scriptstyle <}\over
         {\scriptstyle \sim}\,$}}
\newcommand\gsim{\buildrel > \over \sim}
\def\lambar{\lambda\llap {--}}

\begin{document}

\title*{PULSAR HIGH-ENERGY EMISSION FROM THE POLAR CAP AND SLOT GAP}


\titlerunning{Emission from the Polar Cap and Slot Gap}

\author{Alice K. Harding\inst{1}}


\institute{
     Astrophysics Science Division, NASA Goddard Space Flight Center \texttt{harding@twinkie.gsfc.nasa.gov}
}
  
\maketitle

\begin{abstract}
Forty years after the discovery of rotation-powered pulsars, we still do not understand many 
aspects of their pulsed emission.  In the last few years there have been some fundamental developments 
in acceleration and emission models.  I will review both the basic physics of the models as well as the 
latest developments in understanding the high-energy emission of rotation-powered pulsars, with 
particular emphasis on the polar-cap and slot-gap models.   Special and general relativistic effects play 
important roles in pulsar emission, from inertial frame-dragging near the stellar surface to aberration, 
time-of-flight and retardation of the magnetic field near the light cylinder.  Understanding how these 
effects determine what we observe at different wavelengths is critical to unraveling the emission 
physics.  I will discuss how current and future X-ray and gamma-ray detectors can test the predictions of 
these models.
\end{abstract}

\section{Introduction} \label{sec:intro}


Rotation-powered pulsars are fascinating astrophysical sources and excellent 
laboratories for study of fundamental physics of strong gravity, strong magnetic fields, high
densities and relativity.  The major advantage we have in studying pulsars is that we know
they are rotating neutron stars and that they derive their power from rotational energy loss.
The challenge is then to understand how they convert this source of power into the visible
radiation.  It is generally agreed that this occurs through acceleration of charged particles to 
extremely relativistic energies, using the rotating magnetic field as a unipolar inductor to create very
high electric potentials.  Beyond this fundamental, there is a large divergence of thought on what
comes next: whether the acceleration occurs in the strong field near the neutron star surface or
in the outer magnetosphere near the speed of light cylinder, or even beyond the light cylinder in
the wind zone.  The particle acceleration may well be occurring in all of these regions, either in 
the same pulsar or in pulsars of different ages.

In recent years, there has been much activity both in new detections and in theoretical study
of rotation-powered pulsars.  Multibeam radio surveys at the Parkes Telescope \cite{Manchester2001,Edwards2001} 
have increased the population of known radio pulsars to more than 1700.  
In addition, extended radio observations of supernova remnants and unidentified $\gamma$-ray sources have
discovered a number of young pulsars that are too radio-faint to be detected by surveys \cite{Camilo2004}.
Although pulsed emission at other wavelengths has been detected from only a small fraction of these,
this number is growing as well.  At the present time, there are 7 pulsars with high-confidence
detection of $\gamma$-ray pulsations \cite{Kanbach2006}, about 30 having X-ray pulsations \cite{Kaspi2006} 
and 10 with optical pulsations \cite{Mignami2004}.  

This paper will review both the fundamental physics and latest theoretical developments of
acceleration and radiation in polar cap models.  A complementary paper by Cheng \cite{Cheng2006} reviews acceleration 
and radiation in the outer gap model.  

\section{Acceleration Near the Polar Cap and Beyond} 
\label{sec:Acc}

A magnetic dipole rotating in vacuum will induce an electric field both along and across the 
magnetic field lines \cite{Deutsch1955}.  In the case of a pulsar, with high angular velocity $\Omega$
and surface dipole fields $B_0 \sim 10^{12}$ G, the electric force parallel to the magnetic field just 
above the neutron star surface exceeds the gravitational force by many orders of magnitude.
Vacuum conditions therefore cannot exist outside a pulsar, since charges can be pulled from the
stellar surface \cite{GoldreichJulian1969}.  If the charge density reaches
the Goldreich-Julian value,
\be
\rho _{GJ}  = {{\bf \nabla \cdot E}\over 4\pi} \approx  - {{\bf \Omega \cdot B}\over 2\pi c},
\ee
derived from the condition
\be
{\bf E} =  - {{\bf (\Omega \times r) \times \rm{B}}\over c},
\ee
then the electric field parallel to the magnetic field vanishes.  This is the force-free solution where 
charges and magnetic field corotate with the star.  Corotation must break down near the light cylinder,
$R_L = \Omega/c$, due to particle inertia.  But if vacuum cannot surround a pulsar, neither
can a completely force-free magnetosphere, since in that case no acceleration of charge, currents or
radiation would exist.  A real pulsar must operate somewhere between the two extremes of the vacuum and 
the force-free states, but a self-consistent, global solution has not yet been found.  Global
magnetospheric simulations study how a rotating neutron star
magnetosphere fills with charge from the vacuum state.  The resulting domes and torii of charge that
build up near the pole and equator \cite{Krauss-PolsMichel1985,Petri2002} seem to be 
unstable \cite{Spitkovsky2004}, but the particle-in-cell codes cannot run long enough to reach a stable
magnetospheric configuration.  However, it seems that a near force-free magnetosphere \cite{Spitkovsky2006} cannot result only from charges flowing out of the stellar surface, but requires
some extra source of charge created in the magnetosphere above the surface.  This extra source
of charge is thought to be production of electron-positron pairs by the photons radiated by 
accelerating particles.  The pulsar magnetosphere must somehow be made up of self-consistent 
force-free and non-force-free regions in balance with each other.  
One way to determine the structure of these regions is to 
study the microphysics of the electrodynamics and charge flow at different sites where acceleration
may occur.  One of these sites, the polar cap accelerator, is near the neutron star surface on the open field lines that cross the light cylinder. 

\subsection{Polar cap accelerators}

\begin{figure}[t]
\centerline{\epsfig{file=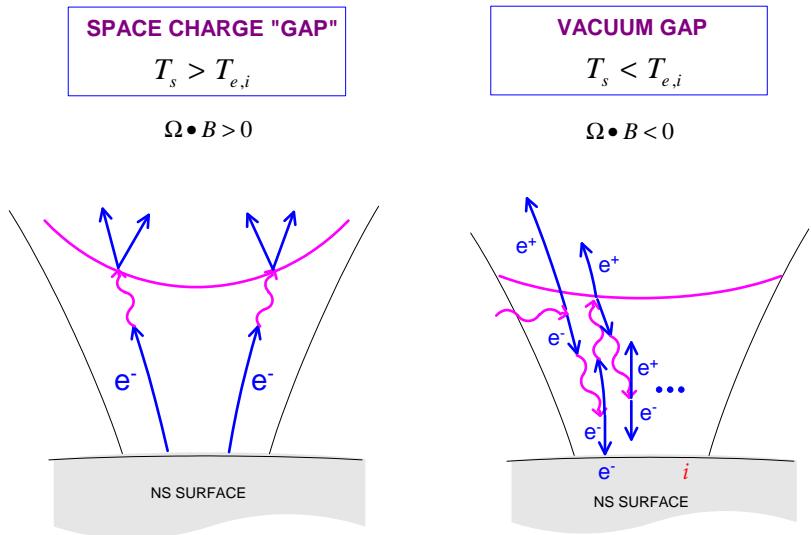,width=12cm}}
\caption{Illustration of space-charge limited flow and vacuum gap accelerators above a pulsar 
polar cap.  $T_s$ is the neutron star surface temperature and $T_{i,e}$ are the ion or electron 
thermionic temperatures.}
\label{fig:1}        
\end{figure}

The two main types of polar cap accelerator are vacuum gaps \cite{RudermanSutherland1975,UsovMelrose1995} and space-charge limited flow (SCLF) gaps \cite{AronsScharlemann1979,HardingMuslimov1998}.  
There are binding (or cohesive) forces on charged particles in the neutron star surface due to the lattice structure in a strong magnetic field, such
that particles are free only if the surface layers are above the thermionic emission temperature.
For electrons, this temperature is \cite{UsovMelrose1995}
\be  \label{eq:Te}
T_e  \cong 3.6 \times 10^5 K^{} \left( {\frac{Z}{{26}}} \right)^{0.8} \left( {\frac{{B_0 }}{{10^{12} G}}} \right)^{0.4} 
\ee
and for ions it is
\be  \label{eq:Ti}
T_i  \cong 3.5 \times 10^5 K^{} \left( {\frac{{B_0 }}{{10^{12} G}}} \right)^{0.73} 
\ee
where $B_0$ is the surface magnetic field strength and $Z$ is the atomic number of matter in the surface layer.
If the surface temperature, $T_s < T_{i,e}$, then charges are trapped in the surface and the full
vacuum electric field, $E_{\parallel} = \Omega B_0 R$,where $R$ is the neutron star radius, 
exists above the surface.  If $T_s > T_{i,e}$,
then charges are ``boiled off" the surface layers and can flow along the open field lines in SCLF.  
Measured surface temperatures of pulsars are typically $T_s > 0.5 - 3.0 \times 10^6$ K, 
above $T_e$ and $T_i$ for the normal range $B_0 \lsim 10^{13}$ of surface fields.  However, a few
high-field pulsars or magnetars may have $T_s < T_{i,e}$ and 
vacuum gaps \cite{UsovMelrose1995,ZhangHarding2000a}.  
If $T_s > T_{i,e}$ and the full Goldreich-Julian charge can be 
supplied right about the surface, the true charge density along each field line will
drop faster,  $\rho \propto r^{-3}$, than $\rho _{GJ}$ as the dipole field lines flare.  
A SCLF electric field,
$\vec{\nabla} \cdot \vec{E_{\parallel}} = (\rho - \rho_{GJ})/\epsilon_0$ grows with altitude.  The
two types of accelerator thus differ by the surface boundary condition: where $\rho (R) = 0$, 
$E_{\parallel}(R) \ne 0$ for vacuum gaps and $\rho (R) = \rho _{GJ}$, $E_{\parallel}(R) = 0$ for SCLF 
accelerators.  The form of $E_{\parallel}$ in
SCLF accelerators is thus sensitive to the detailed distribution
of the charge density, which depends both on the open field line
geometry as well as the compactness of the neutron star.  At
altitudes $\{z \ll \theta_{PC}$, $z \gg \theta_{PC}\}$, with $z
\equiv (r/R - 1)$ being the height above the surface,  
\begin{eqnarray}  \label{eq:Epar} 
& E_{_{||} } \simeq B_0 \theta_{_{PC}}^2 \left[\{{z,
\theta _{_{PC}}^2 \left({r \over R}\right)^{-4}}\} \kappa \cos
\alpha + \{z, \theta _{_{PC}}^2 \left({r \over R}\right)^{-1/2}\}
\frac{\theta _{_{PC}}}{2} \sin \alpha \cos \varphi \right]  \nonumber \\
& \times [1 -\left({\theta \over \theta_{_{PC}}}\right)^2] 
\end{eqnarray}
\cite{MuslimovTsygan1992,HardingMuslimov1998}, where $\theta$ and $\varphi$ are the
magnetic polar and azimuth angles, $\alpha$ is the magnetic inclination angle to the rotation axis, 
$\kappa = 2 G I /(c^2 R^3)$ is the stellar compactness parameter, $\theta_{_{PC}} \simeq (\Omega R/c)^{1/2}$
is the polar cap half-angle and $I$ the neutron star moment
of inertia. The first term in Eqn (\ref{eq:Epar}) is due to inertial
frame dragging near the neutron star surface, and dominates for
small $r$ and low inclination, while the second term is due to the
flaring of the field lines.

Both accelerators will be self-limited by the development of pair cascades, at altitudes where the particles reach high enough Lorentz factors to radiate $\gamma$-ray photons.  The dominant 
process in most pulsars is one-photon pair production, which can occur only in very strong magnetic
(or electric) fields.  In this process, the magnetic field absorbs the extra momentum of a photon 
having the energy required to create
a pair, so that the threshold condition on the photon energy is 
$\epsilon_\gamma = 2mc^2/\sin\theta_{\gamma B}$, where $\theta_{\gamma B}$ is the angle between the 
photon propagation direction and the local magnetic field.  The accelerated particles moving along 
magnetic field lines with high Lorentz factors radiate $\gamma$-ray 
photons at very small angles to the field ($\theta_0 \sim 1/\gamma$), so the one-photon pair production rate 
for these photons is initially zero.  However, as they propagate through the curved dipole field, their angle 
increases until the threshold condition and the attenuation coefficient becomes large.
The voltage across a vacuum gap breaks down 
when a stray $\gamma$-ray crosses the magnetic field within the gap and creates a pair.  The electron
accelerates upward/downward for $\Omega \cdot B (>,<) 0$ over the polar cap, and the positron accelerates in 
the opposite direction as shown in Figure \ref{fig:1}. 
Both particles produce more pairs when their radiated photons reach
the pair threshold, causing a pair avalanche and sudden discharge of the vacuum. The potential
drop in the gap thus oscillates between $V_{vg} \sim \Omega B_0 (R\theta_{_{PC}})^2/2$ and 0. 
In SCLF for $\Omega \cdot B (>,<) 0$, an electron/positron accelerates upward from the surface until the
radiated photons reach pair threshold, at which point the positrons/electrons from the pair decelerate, 
turn around, and
accelerate downward toward the neutron star surface.  The polarization of pairs above the pair formation front
(PFF), discussed in more detail in Section \ref{sec:PCheat}, may short out the $E_{\parallel}$, halting any further acceleration at higher altitude.  These accelerators
can thus maintain a steady current of upwardly accelerating
electrons, at $j^{-}_{\parallel} \simeq c\rho _{GJ}$, and a downward
current of positrons, at $j^{+}_{\parallel} \ll c\rho _{GJ}$,
which heat the polar cap.  The accelerator voltage is determined
by the height of the PFF, which is again roughly comparable to the pair creation mean-free path.
However, the stability of SCLF accelerators has not yet been verified through time-dependent models, and
some simplified studies \cite{Levinson2005} have in fact shown that some oscillations in the pair creation
rate could exist.  This is certainly an issue that needs further investigation. 
Stability of vacuum gaps has also been studied by Gil et al. \cite{Gil2006}.

\subsection{Death lines}
\label{sec:DeathLines}

\begin{figure}[t]
\centerline{\epsfig{file=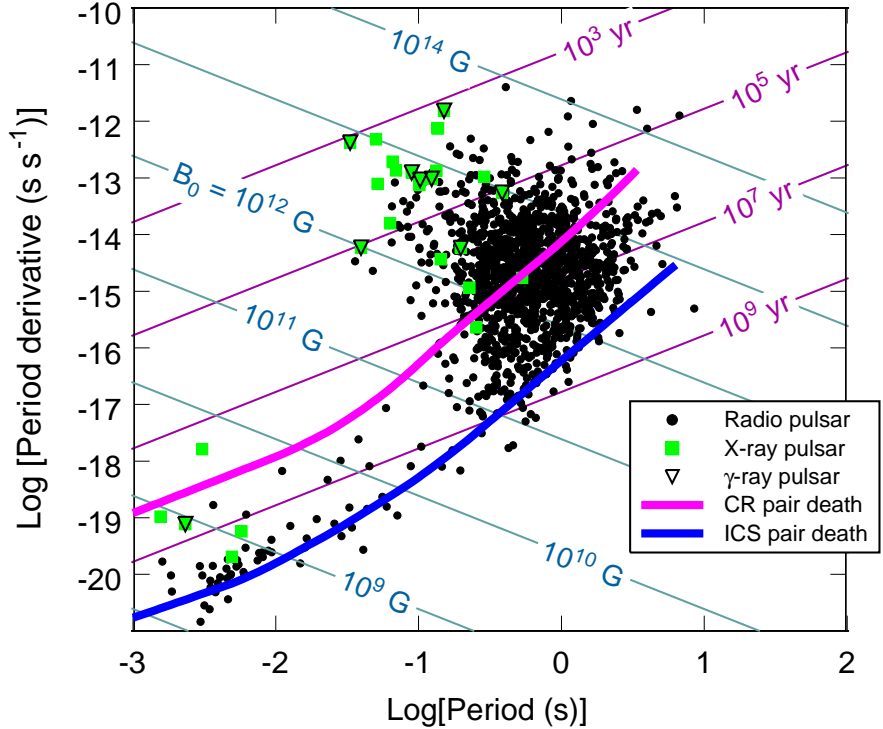,width=13cm}}
\caption{Period derivative vs. period for radio and high-energy pulsars from the ATNF catalog, 
showing death lines \cite{HardingMuslimov2002} for 
production of electron-positron pairs through curvature radiation (CR) and inverse Compton scattering (ICS).
The ICS pair death line was determined for a neutron star surface temperature of $10^6$ K and a standard 
equation of state \cite{Baym1971}.}
\label{fig:2}        
\end{figure}

The pair cascades can be initiated either by curvature radiation (CR) \cite{DaughertyHarding1982} 
or by resonant or non-resonant inverse-Compton scattering (ICS) of stellar thermal X-rays by primary electrons \cite{Sturner1995}.  Since for a given Lorentz factor the peak CR photon 
energy, $\epsilon_\gamma =  3\lambda_C \gamma^3/ 2\rho_c$, where $\lambda_C \equiv h/mc$ is the electron Compton wavelength and $\rho_c$ is the magnetic field radius of curvature, 
is much lower than the ICS peak energy, $\epsilon_\gamma \sim \gamma$ in the extreme
Klein-Nishina limit, pair production of CR photons requires a much higher particle Lorentz factor.
The PFF for ICS therefore occurs at a lower altitude than the PFF for CR \cite{HardingMuslimov1998}.
The PFF altitude is the sum of the acceleration length $\propto 1/E_{\parallel}$ and the pair attenuation 
length, which are both inverse functions of the magnetic field, and effectively of the pulsar age.  
If the PFF is larger than a stellar radius, then the magnetic field becomes too weak for pair creation to occur and a PFF does not exist.
For SCLF accelerators, CR photons can produce pairs only in the case of
young pulsars ($\tau \lsim 10^7$ yr) and a few millisecond pulsars  
\cite{HardingMuslimov2001,HibschmanArons2001}.  
One can define a `death line' in $\dot P - P$ space, where $\dot P$ is the period
derivative, shown in Figure \ref{fig:2} below which pulsars cannot produce 
pairs from CR photons.  Below the CR pair death line, pulsars can 
produce pairs only from ICS photons \cite{HardingMuslimov2002}.  
Below a lower, ICS pair death line, pulsars cannot produce any pairs and are expected to be radio quiet.  
In fact the predicted ICS death line falls close to the edge of the known radio pulsar population, as shown 
in Figure \ref{fig:2} \cite{Harding2002}.  The few millisecond pulsars that lie below the ICS pair death 
line may be able to produce pairs through interaction of ICS photons with thermal X-ray photons from 
a hot polar cap \cite{ZhangQiao1998}.
 
\section{Electric Field Screening and Polar Cap Heating}  
\label{sec:PCheat}

\begin{figure}[t]
\centerline{\epsfig{file=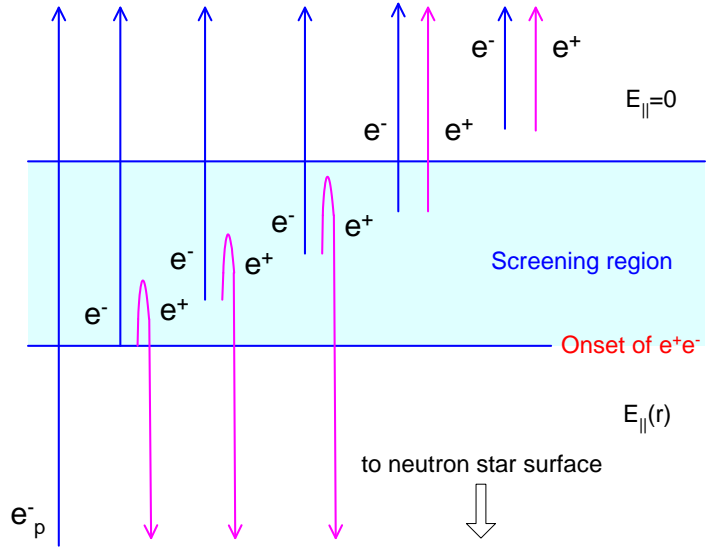,width=11cm}}
\caption{Illustration of electric field screening in SCLF models.  $e_p^-$ represents the primary
electron flux accelerating upward from the neutron star surface.  $e^+$ and $e^-$ represent electron-positron 
pairs created by the primary electron above the pair formation front.}
\label{fig:3}        
\end{figure}

In SCLF accelerators, the polarization of charge above the PFF acts both to screen the $E_{\parallel}$
and to produce heating of the polar cap by the downward flowing particles \cite{Arons1981}.  
Figure \ref{fig:3} illustrates
the dynamics of electric field screening.  Primary electrons ($e_p^-$) accelerate upward from the stellar surface and 
produce pairs at different altitudes above the PFF.  The positrons decelerate and turn around in a distance 
short compared to the PFF altitude and each reversing positron creates a small excess of negative charge.
As more positrons are produced and decelerated, the space charge becomes more negative until the entire
charge deficit  $\delta \rho = (\rho - \rho_{GJ})$ that produced the $E_{\parallel}$ is accounted for.  
Since the 
charge deficit is small compared to the primary charge ($\delta \rho \ll \rho_{GJ})$, the screening length
scale is a very small fraction of the PFF altitude.  The flux of returning positrons, as a fraction of the 
primary flux, is approximately
\be  \label{f+}
f_+  \approx \frac{{\rho _ +  }}{{\rho _{GJ} }} = \left. {\frac{{\rho _{GJ}  - \rho }}{{2\rho _{GJ} }}} \right|_{z_0 }  \approx \frac{3}{2}\frac{\kappa }{{(1 - \kappa )}}z_0 
\ee
where $z_0$ is the PFF height above the stellar surface and $\kappa$ is the general relativistic factor appearing in the SCLF electric field (Eqn(\ref{eq:Epar})).  
The corresponding polar cap heating luminosity is 
\be  \label{L+}
L_ + ^{\max }  \approx f_ +  \Phi (z_0 )\mathop {\dot n_{\rm prim} } 
\ee
where $\dot n_{\rm prim}$ is the primary particle flux and $\Phi (z_0 )$ is the potential drop at $z_0$.  
The growth in the charge density above the CR PFF is
very rapid since the CR peak energy $\epsilon \propto \gamma^3$ increases rapidly with increasing electron Lorentz factor, $\gamma$.  The CR initiated cascades have very high multiplicities and the screening of 
$E_{\parallel}$ takes place in a relatively short distance above the PFF \cite{HardingMuslimov2001}.  
The ICS initiated cascades on the
other hand have much lower multiplicities since the ICS photon production rate decreases with 
increasing $\gamma$, as $\sim 1/\gamma$, for non-resonant scattering and first increases sharply, then 
decreases as $\sim 1/\gamma^2$ for resonant scattering.
The screening above ICS PFFs takes place over a larger scale length, depending on the polar cap
temperature and for a number of pulsars well below the CR pair death line, there is no screening \cite{HardingMuslimov2002}.  Even when
ICS screening is locally complete, an unscreened charge deficit can develop at higher altitudes since
the charge density deficit $\delta \rho$ grows faster than the increase in pair density.  
In that case, an $E_{\parallel}$ reappears and acceleration continues.  If a CR PFF forms at higher 
altitude, then screening occurs, otherwise unscreened acceleration continues to high altitude 
\cite{MuslimovHarding2004b}.

\begin{figure} [t]
\centerline{\epsfig{file=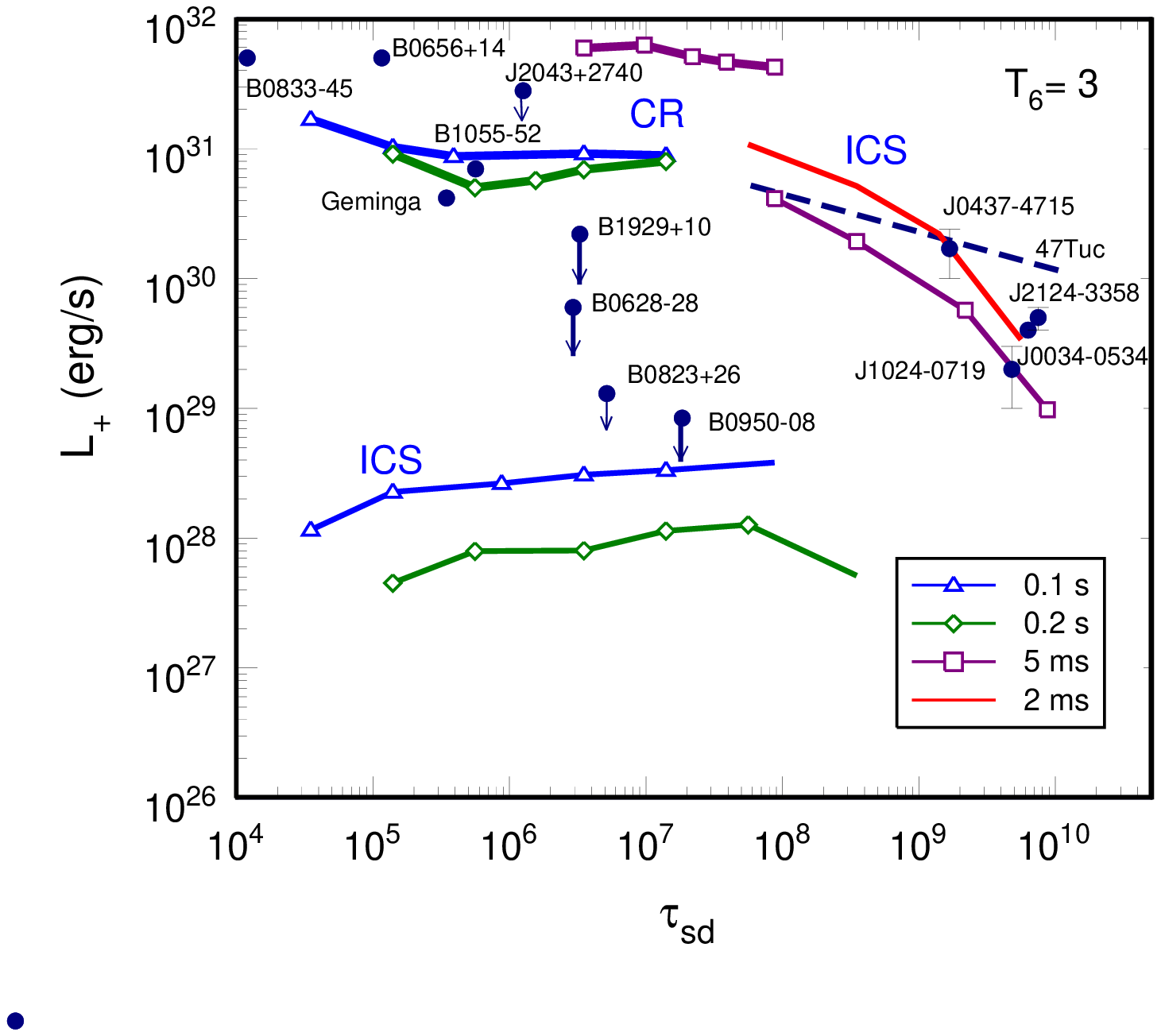,width=13cm}}
\caption{Predicted polar cap heating luminosity, $L_+$, for SCLF model vs. spin down age \cite{HardingMuslimov2002}, $\tau_{sd}$, for 
different pulsar periods and an assumed polar cap surface temperature of $3 \times 10^6$ K.  Thick curves
are luminosities from curvature radiation (CR) pair heating and thin lines are luminosities from inverse Compton
(ICS) pair heating.  The solid circles show measured pulsar luminosities and upper limits of hot thermal 
components (see \cite{Kaspi2006}).  
The dashed line is the fit of measured luminosity vs. age for the millisecond pulsars
in the globular cluster 47 Tuc \cite{Grindlay2002, Bogdanov2006}.}
\label{fig:4}        
\end{figure}

The CR pair heating luminosity is much higher than the ICS heating luminosity because the CR PFF occurs at
higher altitude and a larger flux of positrons returns through a higher voltage, bombarding the polar cap with higher energy.  Figure \ref{fig:4} shows calculated heating luminosities from CR and ICS positrons as a function
of pulsar characteristic age, $\tau_{\rm sd} = P/2\dot P$, 
for several different periods and a PC surface temperature of $T = 3 \times 10^6$ K. 
The CR heating lines terminate on the high $\tau_{\rm sd}$ side at the age corresponding to the CR pair death line, which is
around $\tau_{sd} \sim 10^7$ yr for periods $P = 0.1, 0.2$ s and around $\tau_{sd} \sim 10^8$ yr for periods 
$P = 5$ ms.  The CR heating luminosities are several orders of magnitude higher than the ICS heating luminosities,
for normal pulsars.  All the $L_+$ are higher for millisecond periods, because the PFFs occur at higher
altitude and the gap voltage is larger.  A sudden drop in $L_+$ is therefore predicted at the 
CR death line.  Pulsed X-ray emission has now been detected from many middle-aged and older pulsars 
\cite{Becker2006}.  
Several
thermal components are often seen, one having lower T and larger area, which may be full surface cooling, and another having a higher T and smaller area, which could be due to polar cap heating.
The luminosity of the hot thermal components, which are shown in Figure \ref{fig:4}, 
may indicate that the heating 
luminosity drops suddenly around $\tau_{sd} \sim 10^7$ yr or somewhat sooner.
To within an order of magnitude, the observed luminosities of the hot thermal components agree with the 
predicted $L_+$, for both normal and millisecond pulsars.  

Although the $L_+$ plotted in Figure \ref{fig:4} were computed numerically, analytic expressions for $L_+$
provide a good estimate in the case of complete screening and upper limits in the case of incomplete
screening.  The heating from CR pair fronts predicts a surface X-ray luminosity of approximately 
\cite{HardingMuslimov2001}
\be \label{L+CR}
L_+^{(CR)} \simeq 10 ^{31}~{\rm erg\, s^{-1}}~  
\left\{ \begin{array}{ll}
    0.4~P^{-6/7}\tau _6^{-1/7} & {\rm if}\: P\lsim 0.1~B_{0,12}^{4/9}, \\
    1.0~P^{-1/2} & {\rm if}~P\gsim 0.1~B_{0,12}^{4/9},
\end{array} 
\right.
\ee
where $B_{0,12} = B_0/10^{12}$ G.  The heating from ICS pair fronts predicts a surface X-ray luminosity of 
approximately \cite{HardingMuslimov2002}
\be
\label{L+ICS}
L_+^{(IC)} \simeq 2.5 \times 10 ^{27}~{\rm erg\, s^{-1}}~P^{-3/2}.
\ee
Since nearly all millisecond pulsars produce only ICS pairs with incomplete screening, the above expression
overestimates the predicted $L_+$ in these sources.

\section{Slot gap accelerator}

Due to the geometry of the field lines and the assumed boundary conditions of the accelerator, the 
altitude of the PFF varies with magnetic colatitude across the polar cap 
\cite{Arons1983,HardingMuslimov1998}.  On field lines well inside the polar cap rim,
$E_{\parallel}$ is relatively strong and the PFF is
very near the neutron star surface.  But at the polar cap rim,
which is assumed to be a perfectly conducting boundary,
$E_{\parallel}$ vanishes.  Near this boundary, the electric field
is decreasing and a larger distance is required for the electrons
to accelerate to the Lorentz factor needed to radiate photons
energetic enough to produce pairs.  The PFF thus 
curves upward as the boundary is approached, forming a narrow slot
gap (see Figure \ref{fig:6}) near the last open field line \cite{Arons1983}.  Since
$E_{\parallel}$ is unscreened in the slot gap, particles continue
to accelerate and radiate to high altitude along the last open
field lines.  The width of the slot gap is a function $\Lambda \equiv P\,B_{0,12}^{-4/7}$ 
of pulsar period 
and surface magnetic field \cite{MuslimovHarding2003}, 
and can be expressed in magnetic colatitude as a fraction of the polar cap
half-angle $\Delta \xi_{\rm SG}$, where $\xi \equiv \theta/ \theta_{\rm PC}$
\be
\Delta \xi_{\rm SG} \simeq 
\left\{ \begin{array}{ll}
 4 \,\Lambda, & ~~~~ \Lambda < 0.075 \\
 0.3,  & ~~~~ \Lambda > 0.075 
\end{array}
\right.
\ee
The particles can achieve very high Lorentz factors which at altitudes of
several stellar radii are limited by curvature radiation losses, to $\gamma_{SG} \simeq 
3-4 \times 10^7$ \cite{MuslimovHarding2004a}.  Since the slot gap is very narrow for young
pulsars having short periods and high fields, the corresponding solid angle of the gap emission 
$\Omega _{SG}  \propto \theta _{PC}^2 r^{} \Delta \xi _{SG}$
is quite small.  So even though only a small fraction of the polar cap flux is accelerated in the
slot gap, the radiated flux $\Phi_{\rm SG} = L_{SG}/\Omega _{SG}\,d^2$ can be substantial.  
The total luminosity divided by solid angle from each pole is 
(from \cite{MuslimovHarding2003})
\begin{eqnarray}  \label{eq:L_SG}
 \frac{L_{SG} }{\Omega _{SG}} = 
\varepsilon _\gamma \,\, [0.123\cos ^2 \alpha  + 0.51\,\theta _{PC}^2 \sin ^2 \alpha ]\,\, {\rm erg\,s^{-1}\,sr^{-1}}
\nonumber \\
\times \left\{
\begin{array}{ll}
9 \times 10^{34}\, L_{sd,35}^{3/7} P_{0.1}^{5/7}, & ~~~ B < 0.1\,B_{\rm cr} \\
2 \times 10^{34}\, L_{sd,35}^{4/7} P_{0.1}^{9/7}, & ~~~B > 0.1\,B_{\rm cr} 
\end{array}
\right.
\end{eqnarray}
where $P_{0.1} \equiv P/0.1$ s, $L_{sd,35} \equiv L_{sd}/10^{35}\,\rm erg\,s^{-1}$  
is the spin down luminosity, $B_{\rm cr} \equiv 4.4 \times 10^{13}$ G is the critical magnetic field
strength and $\varepsilon _\gamma$ is 
the efficiency of conversion of primary particle energy to high-energy emission.

For $\Lambda \gg 0.075$, which corresponds to pulsars below the CR pair death line (see Fig. \ref{fig:2}), 
the slot gap disappears since the screening of $E_{\parallel}$ is no longer effective.  These 
``pair-starved" pulsars may have local screening of $E_{\parallel}$ near the ICS PFF, but $E_{\parallel} \ne 0$
at higher altitude so that acceleration can occur over nearly all of the open field volume 
\cite{MuslimovHarding2004b}.  The particle Lorentz factors, as in the slot gap, will be limited by 
curvature-radiation reaction to $\gamma \sim 10^7$.
 
\section{High-Energy Radiation}

\subsection{Polar cap and slot gap cascades}

As discussed in Section \ref{sec:DeathLines}, pair cascades above the polar cap may be initiated by either
ICS or CR.  In the strong magnetic fields near the neutron star, the Compton scattering cross section has a
resonance at the cyclotron energy, $\epsilon_B \sim 12 B_{12}$ keV, where its value is several orders of
magnitude higher than the Thompson cross section.  The thermal X-rays from the neutron star surface will be
blueshifted into the resonance in the rest frame of the accelerating primary electrons when $kT\gamma (1-\beta\cos\theta) \simeq \epsilon_B$, where $\theta$ is the angle between the photon propagation and the 
electron velocity (i.e. the magnetic field direction).  Since the enhancement of the cross section at the
resonance is so large, resonance scattering will dominate if the resonance condition is met 
\cite{DaughertyHarding1989, Xia1985}.  For typical pulsar fields and surface temperatures, electrons with 
$\gamma \sim 10^2 - 10^6$ will resonant scatter surface thermal X-rays 
\cite{ZhangQiao1996,HardingMuslimov1998} 
to energies $\epsilon \simeq \gamma B'mc^2$, where $B' \equiv B/B_{\rm cr}$ is the magnetic 
field strength in units of the critical field strength, $B_{\rm cr} = 4.4 \times 10^{13}$ G.   X rays
scattered by electrons with $\gamma \sim 10^5 - 10^6$ will produce pairs to form a PFF.  Resonant
scattering is the dominant mode of ICS pair production for pulsars having the highest fields 
\cite{HibschmanArons2001}, whereas pulsars with lower fields, that include the millisecond pulsars, produce ICS
pairs through non-resonant scattering.  For $\gamma \gsim 10^5$, the electrons will scatter non-resonantly 
in the Klein-Nishina limit, to energies $\epsilon \simeq \gamma$.  The ICS pair cascade multiplicities are
$M_+^{ICS} \simeq N_+/N_p \sim 10^{-3} - 10$ (pairs per primary electron) \cite{HibschmanArons2001}, 
which is too low and produced at too low an 
altitude to completely shut off the acceleration.  When the primary electrons reach $\gamma \gsim (0.5 - 1) 
\times 10^7$, they produce CR photons that can produce pairs to form a PFF at height $0.02 - 0.1$ stellar
radii.  The CR pair cascade multiplicities reach as high as $M_+^{CR} \simeq 10^3 - 10^4$ 
\cite{HibschmanArons2001, ArendtEilek2002}.  

The pairs are produced in 
Landau states whose maximum principal quantum number is $n_{\rm max} = 2\epsilon'(\epsilon' - 2)/B'$ 
\cite{DaughertyHarding1983}, where $\epsilon' = \epsilon\sin\theta$ is the photon energy in the frame in
which it propagates perpendicular to the local magnetic field.  When the local field $B \lsim 0.1 B_{\rm cr}$, pairs will be created above threshold in highly excited
Landau states and the excitation level is quite sensitive to field strength, $n_{\rm max} \propto 1/B'^3$.  
The pairs will decay through
emission of synchrotron/cyclotron photons, many of which will produce more pairs in excited states.  A
pair cascade can be sustained in such a way through several generations.  The pairs can also produce ICS photons, through scattering surface thermal X-rays, which may produce pairs
\cite{ZhangHarding2000b}.  In high-field pulsars,
where $B \gsim 0.1 B_{\rm cr}$ near the neutron star surface, the pair creation attenuation coefficient 
is high enough for pair creation near threshold, so pairs are produced in very low Landau states.  In this 
case, cascade pair multiplicities are lower since the number of cyclotron photons drop significantly 
\cite{BaringHarding2001}.  Bound pair production \cite{ShabadUsov1982} also becomes important
for $B \gsim 0.1 B_{\rm cr}$, which will further lower the pair multiplicity.  
When $B \gsim B_{\rm cr}$, photon splitting dominates over pair production \cite{Harding1997}.  
The screening of the electric field is also delayed by splitting and 
creation of bound pairs, effectively increasing the accelerating potential.  The effect of
bound pair creation on polar cap acceleration has been studied by Usov \& Melrose \cite{UsovMelrose1996}.  

Radiated spectra from CR-initiated \cite{DaughertyHarding1982} and ICS-initiated 
\cite{Sturner1995} cascades 
are very hard (roughly power laws with indices 1.5 - 2.0) 
\cite{HardingDaugherty1998} with sharp cutoffs due
to magnetic pair production at energy \cite{Harding1997, Harding2001}
\be
\label{Ec}
E_c \sim 2\,\,{\rm GeV}\,P^{1/2}\,\left({r\over R}\right)^{1/2}\, {\rm max}
\left\{0.1, \,B_{0,12}^{-1}\,\left({r\over R}\right)^3\right\}
\ee
where $r$ is the emission radius and $R$ is the neutron star radius.  An approximate form for the
polar cap pair cascade spectrum is given by
\be
\label{eq:PCcaspec}
f(\epsilon) = A \epsilon^{-a}\,\exp\left[{-C_{1\gamma}\exp\left(-{\epsilon^{1\gamma}_{\rm esc}/\epsilon}\right)}\right]
\ee
where $a$ is the power-law index, $C_{1\gamma} = 0.2(\alpha/\lambar)(B'R^2/\rho)$ and $\lambar = \lambda_C/2\pi$ 
is the electron Compton wavelength.
The polar cap cascade radiation produces a hollow cone of emission around the magnetic pole, with opening angle, $\theta_{\gamma} \simeq (3/2)(\Omega r/c)^{1/2}$, determined by the polar cap half-angle at the radius of emission $r$.  The characteristics of emission from this
type of polar cap model \cite{DaughertyHarding1996} has been successful in reproducing many 
features of $\gamma$-ray pulsars, including the wide double-peaked pulses observed from $\gamma$-ray pulsars 
like the Crab, Vela and Geminga, and the phase-revolved spectra.  
However, the polar cap opening angle is very small (a few degrees) unless the emission occurs more than a few 
stellar radii above the surface \cite{DyksRudak2002}.  Since the pair cascades over most of the polar cap occur 
within several stellar radii of the stellar surface, the broad profiles which require beam opening 
angles of the order of the magnetic inclination angle $\alpha$, cannot be produced unless the pulsar is 
nearly aligned.  Daugherty \& Harding \cite{DaughertyHarding1996} 
had to assume extended acceleration to 3 stellar radii and an artificial  
enhancement of primary particle flux near the polar cap rim in order to reproduce the Vela pulsar spectrum
and pulse profile.

\begin{figure}[t]
\centerline{\epsfig{file=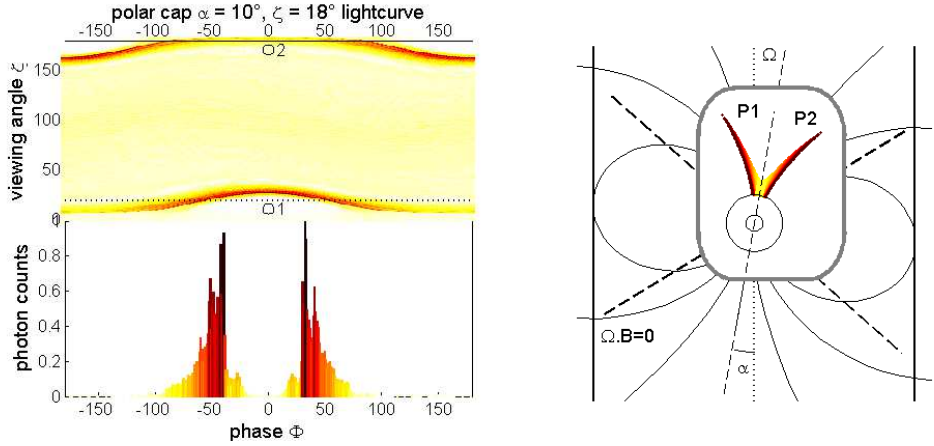,width=13cm}}
\caption{Phase plot, sample lightcurve, and a sketch of the
accelerator location for the polar cap model, for a typical
inclination angle $\alpha = 10^\circ$. The central zoom gives the
gap extent relative to the star size. The dashed lines outline the
null surface. The shading in the lightcurve and gap sketch is the
same. The phase plot illustrates the change in lightcurve as seen
by different observers and the aperture of the pulsed beams. From \cite{GrenierHarding2006}.}
\label{fig:5}        
\end{figure}

Muslimov \& Harding \cite{MuslimovHarding2003} found that pair cascades on the inner edge of the slot gap occur at altitudes of 3-4 stellar radii and 
have higher multiplicities, $M_+ \sim 10^4 - 10^5$, than the polar cap cascades.  These are the characteristics
required by \cite{DaughertyHarding1996} 
to model $\gamma$-ray pulsars, so a combination of polar cap cascades near the stellar
surface and slot gap cascades near the polar cap rim at higher altitude may be successful in explaining
non-thermal radiation from some fraction of sources having small magnetic inclination $\alpha$ and 
viewing angles $\zeta$.  Figure \ref{fig:5} shows a plot of the total sky emission, phase $\Phi$ vs. viewing
angle $\zeta$ relative to the rotation axis, for a slot gap cascade at inclination angle 
$\alpha = 10^\circ$.  In this model, the magnetic pole is at phase $0^\circ$, midway between the two peaks
in the profile, which would be the predicted phase peak of polar cap thermal emission. 

\subsection{Radiation from the high-altitude slot gap}
\label{sec:SGrad}

The electrons that accelerate in the slot gap and generate pair cascades at low altitude continue to
accelerate.  They will radiate curvature, inverse Compton and synchrotron radiation
at high altitudes.  Initially, their Lorentz factors will be limited by curvature-radiation reaction, to
\be  \label{eq:gamCRR}
\gamma _{_{CRR}}  = \left( {\frac{3}{2}\frac{{E_{||} \rho _c^2 }}{e}} \right)^{1/4}  \sim 3 \times 10^7 
\ee
and the peak energy of their CR spectrum will be $\varepsilon _{peak}^{CR}  = 2\lambda _C \gamma _{_{CRR} }^3 /\rho _c  \approx 3{\rm{0 GeV}}$.  As the electrons reach higher altitude, where the local magnetic field
has dropped to $B \sim 10^6 - 10^8$ G, they may be able to resonantly absorb radio photons of energy $\varepsilon _{0,GHz}$ that are at the cyclotron resonance in their rest frame .  
The condition for resonant absorption is
\be  \label{gammaR}
\gamma_R  = 3 \times 10^5 \frac{{B_8 }}{{\varepsilon _{0,GHz} (1 - \beta \mu _0 )}}
\ee
where $B_8 \equiv B/10^8$ G and $\mu_0$ is the cosine of the angle between the radio photon and the electron
momentum.  The electrons are then excited to higher Landau states and radiate synchrotron emission. 
Blandford \& Scharlemann \cite{BlandfordScharlemann1976} investigated this process for pulsars, concluding that it was not
an important source of radiation.  However, they assumed that the electrons would be excited to only low Landau states and based their estimate on the cyclotron emission rate.  Decades later, Lyubarsky \& Petrova \cite{LyubarskyPetrova1998}
re-examined resonant cyclotron absorption of radio emission in pulsar magnetospheres and discovered that
the rate of absorption is much higher than the rate of re-emission in low-lying Landau states, so the
electrons will continue to be excited to high Landau states, and their pitch angles will increase until
their momentum perpendicular to the magnetic field is relativistic.  The emission is then synchrotron, not
cyclotron, which proceeds at a much higher rate.  Petrova \cite{Petrova2003} concluded that this process could be a 
significant source of radiation for young pulsars.  It could also be significant for millisecond pulsars with
high radio luminosities like PSR J0218+4232 (\cite{Harding2005} and Section \ref{MSP}).  

In the slot gap or along open field line of pair-starved pulsars, electrons can experience both continuous
acceleration and resonant cyclotron absorption.  A steady-state may be reached between synchrotron radiation
(SR) losses and absorption, with perpendicular momentum
\be
{p_ \bot ^{SRR}\over mc}  \simeq 302^{} B_8^{ - 1} E_{\parallel ,5}^{1/2} 
\ee
which can indeed be relativistic.  In this state, the Lorentz factor can remain locked at $\gamma_R \ll 
\gamma_{_{CRR}}$ to the
cyclotron resonance as the electron moves along the field with the field decreasing \cite{Harding2005}.
The synchrotron peak frequency decreases with increasing altitude, spreading the emission over a range of
energies up to $\sim 100$ MeV, and the curvature radiation rate drops dramatically. 

Non-resonant inverse Compton scattering of the radio emission by relativistic electrons in the slot gap is also possible.  The scattering would be completely in the Thompson regime, and the spectrum would extend to a maximum 
energy of $\varepsilon _{\max }  \sim \gamma _{CRR}^2 \varepsilon _{0,GHz} \sim$ a few GeV.  This process might
make a significant contribution to the total spectrum of the slot gap if the radio emission region is located
at relatively high altitude (see Section \ref{sec:radio}).

\subsection{Relativity, geometry and caustics}
\label{sec:caustic}

\begin{figure}[t]
\centerline{\epsfig{file=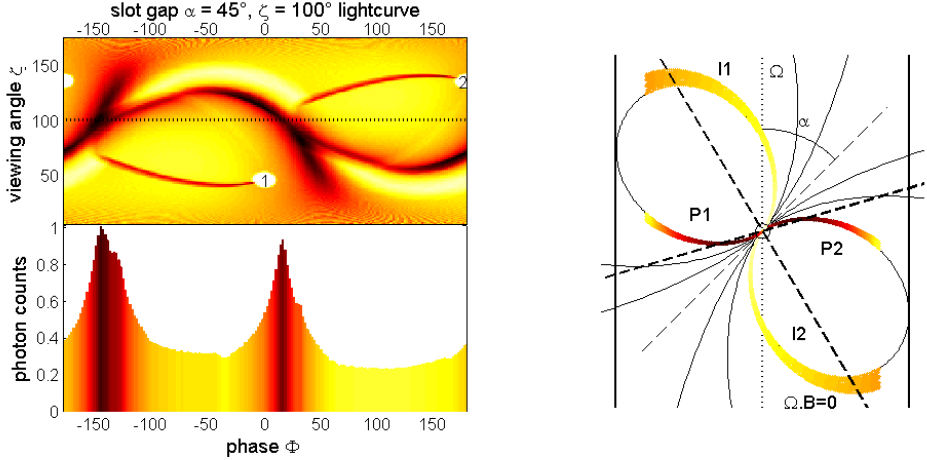,width=13cm}}
\caption{Same as Figure \ref{fig:5} for the slot gap model, for a
typical inclination angle $\alpha = 45^\circ$. From \cite{GrenierHarding2006}.}
\label{fig:6}        
\end{figure}

The geometry of emission at high altitude is strongly influenced by special relativistic effects of aberration, time-of-flight and retardation of the magnetic field.  Aberration and time-of-flight 
produce phase shifts of comparable magnitudes $\Delta \Phi \sim -r/R_{LC}$ in radiation emitted at different altitudes $r$ relative to the light cylinder radius $R_{LC}$.  Morini \cite{Morini1983} first noted that the combined
phase shifts from aberration and time-of-flight, of photons radiated tangent to a magnetic dipole field 
from the polar cap to the light cylinder, nearly cancel those due to field line curvature on  
the trailing edge of the open-field region.  Radiation along such trailing field lines 
bunches in phase, forming a sharp emission peak or caustic in the phase plot or profile (Figure \ref{fig:6}).
On the leading side, these phase shifts add up to spread photons emitted at various
altitudes over a large range of phases.  The effect is most pronounced for large $\alpha$ and emission
between altitudes $r_{\rm em} \sim 0.2 - 0.8\,R_{LC}$.  
Sweepback of the magnetic dipole
field due to retardation \cite{Deutsch1955, Yadigaroglu1997} affects photon emission directions near the
light cylinder, and also distorts the polar cap and open field volume \cite{ArendtEilek1998, 
DyksHarding2004}, even at the stellar surface.  Most observer angles sweep across caustics from
both magnetic poles, resulting in a double peaked pulse profile where the peaks generally have phase
separation less than $180^\circ$ \cite{DyksRudak2003}.  Furthermore, emission occurs at all phases in 
the profile.  Such profiles are very similar to those of the bright $\gamma$-ray pulsars, Crab and Vela.
The predicted polarization characteristics of such a ``two-pole caustic" model can also explain the
observed optical polarization of the Crab pulsar \cite{Dyks2004}.

Radiation from the slot gap has a geometry very similar to that of the ``two-pole caustic" model
and thus displays caustics in the intensity phase plots as well as Crab-like pulse profiles 
\cite{MuslimovHarding2004a}. 
The high-altitude slot gap thus may be a viable model for high-energy emission from young pulsars.
However particle acceleration in the slot gap may not operate at all inclination
angles.  The low altitude $E_{\parallel}$, shown in Eqn (\ref{eq:Epar}), is a function of both $\alpha$
and $\phi$.  For large $\alpha$ and $\cos\phi < 1$ (field lines curving away from the rotation axis),
the second term in Eqn (\ref{eq:Epar}) can become large and negative at large $r$, causing the 
$E_{\parallel}$ to reverse sign.  This would cause a buildup of charge at that location, and would clearly 
be an unstable situation.  The charge flow along those field lines might then be either time-dependent
or non-existent.  The solution for $E_{\parallel}$ at high-altitude in the slot gap \cite{MuslimovHarding2004a} may moderate
this effect somewhat, but for fast rotators and $\alpha \gsim 70^\circ$ the $E_{\parallel}$ reverses sign
at low altitude.  The resolution to this problem is still forthcoming, but it is possible that steady
current flow in the slot gap does not occur for all geometries.

\subsection{Radiation from millisecond pulsars}
\label{MSP}

\begin{figure}[t]
\centerline{\epsfig{file=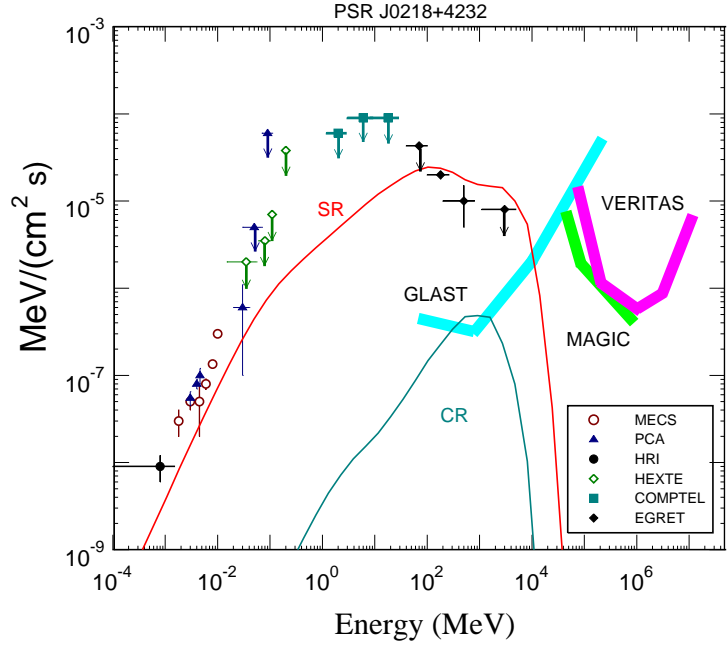,width=12cm}}
\caption{Model spectrum for 2.3 ms PSR J0218+4232 (from \cite{Harding2005}) 
showing components from synchrotron radiation (SR) and
curvature radiation (CR) from unscreened acceleration of a single primary electron, from the
NS surface to the light cylinder, along a field line defined by magnetic colatitude $\xi = 0.9$ in units
of PC half angle, at inclination $\alpha = 50^\circ$.  Data points are from \cite{Kuiper2003}.}
\label{fig:7}        
\end{figure}

There are more than 200 radio pulsars now known with periods between 1 and 30 milliseconds.  Many
(around 30) have been detected as X-ray point sources, and seven of these have X-ray pulsations.
Only the few millisecond pulsars (MSPs) that lie above or near the CR pair death line are expected to 
have slot gaps.  These include B1821-24, B1957+20 and B1937+21, whose pulse profiles interestingly 
show caustic-like narrow, sharp peaks.  The rest that lie below the CR death line are pair-starved and
their SCLF $E_{\parallel}$ is unscreened.  Particles on all open field lines will therefore accelerate
to high-altitude with Lorentz factors limited by CR losses, as in the slot gap (Eqn [\ref{eq:gamCRR}]).
The peak energy of the CR spectrum for MSPs is similarly high, 
$\varepsilon _{peak}^{CR}  \approx 1{\rm{0 GeV }}B_8^{3/4} P_{ms}^{ - 5/4} \kappa _{0.15}^{3/4}$, where
$P_{ms} = P/1 ms$ and $B_8 = B/10^8$ G.  But the very low magnetic fields of MSPs do not absorb the
high energy part of the spectrum, so these sources may be visible up to energies of 50 GeV 
\cite{Bulik2000, Luo2000, Harding2002}.  The CR will dominate for
viewing angles near the magnetic poles because the emission comes primarily from low altitude, since the particle Lorentz factors decrease with increasing altitude.  Additionally,
as discussed in Section \ref{sec:SGrad}, synchrotron radiation from resonant absorption of radio emission
may be important for MSPs having high radio luminosities.  The synchrotron component will appear in the
X-ray to $\gamma$-ray ($\lsim 100$ MeV) region of the spectrum \cite{Harding2005}, and will be visible
at a larger range of observer angles since the emission comes from high altitudes.   Figure \ref{fig:7} shows a model spectrum for the millisecond pulsar PSR J0218+4232 based on a (1D) cyclotron resonant absorption model described above (see Section \ref{sec:SGrad}).

\section{Pulsar Emission at Multiwavelengths}
\label{sec:MWave}

\begin{figure}[t]
\centerline{\epsfig{file=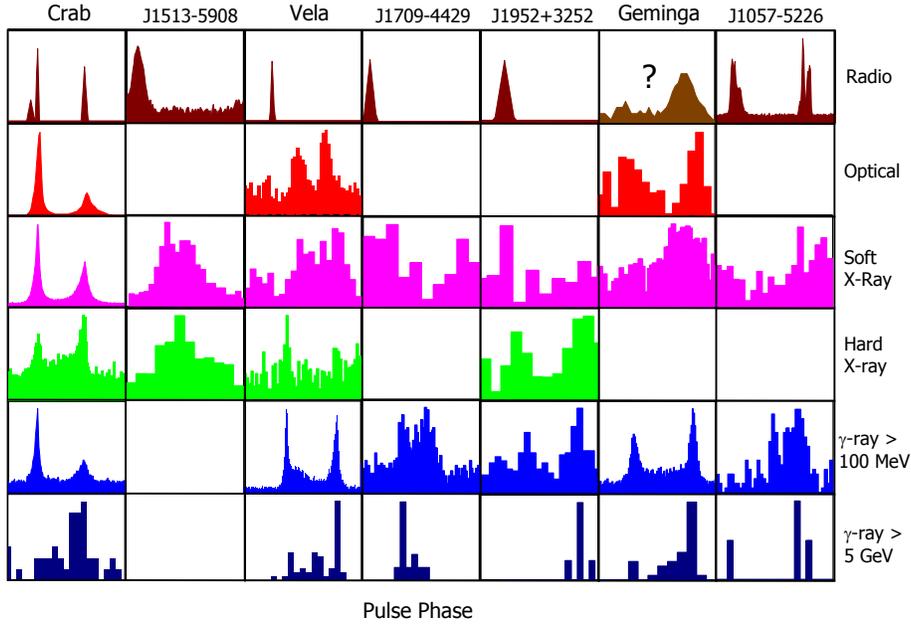,width=13cm}}
\caption{Profiles of seven $\gamma$-ray pulsars in six energy bands, 
as in \cite{Thompson2004}.}
\label{fig:8}        
\end{figure}

Since rotation-powered pulsars shine over a broad spectrum from radio to high-energy $\gamma$-ray wavelengths,
the multiwavelength spectra and profiles can give important clues to the acceleration and emission geometry.
Although there are some clear patterns of spectral behavior with pulsar age \cite{Becker2006},
the observed multiwavelength lightcurves show a wide variety of characteristics that present many puzzles.
As shown in Figure \ref{fig:8}, the lightcurves for the known $\gamma$-ray pulsars do not as a rule exhibit
much phase correspondence between radio, X-ray and $\gamma$-ray peaks.  The sole example is the Crab pulsar, where
two peaks separated by $\sim 140^\circ$ are in phase from radio to $\gamma$-ray energy.  For the other   
pulsars, the radio peak usually leads the one or two $\gamma$ peaks in phase and the soft 
X-ray peaks are broader and 
overlap the $\gamma$-ray peaks.  Comparing X-ray and radio profiles of known rotation-powered 
X-ray pulsars, one sees that there is more multiwavelength phase coherence for shorter periods ($P \lsim 50$ ms),
such as for PSR B0540-69 (50 ms), PSR J1617-5055 (69 ms) and the millisecond pulsars, PSR B1821-24, PSR B1937+21
and PSR J0218+4232.  Such short-period and younger pulsars also tend to have pure power law X-ray spectra with 
no thermal components \cite{Becker2006}, indicating that their X-ray emission is primarily non-thermal radiation 
from accelerated particles.  One possible picture is that 
pulsars with shorter periods accelerate particles mostly in
the outer magnetosphere, in slot gaps or outer gaps.  The high-energy emission occurs at high altitudes that are
large fractions of their light cylinder radii, $R_{LC} = 5 \times 10^{9} P^{-1}$ cm, which are also smaller.  
Their emission geometry is therefore expected to be different from that of longer period pulsars, 
whose high-energy emission occurs closer to the neutron star surface and at altitudes that are very small 
fractions of $R_{LC}$.

\subsection{Radio emission geometry}
\label{sec:radio}

Because the mechanism responsible for the radio beams is not understood, and more importantly 
because the radiation is coherent, it has not been possible to describe this emission using a physical 
model.  The emission has therefore been described using empirical models, developed over the years 
through detailed study of pulse morphology and polarization characteristics.  
The emission is also highly polarized, and displays changes in polarization position angle across the 
profile that often matches the swing expected for a sweep across 
the open field lines near the magnetic poles in the Rotating Vector Model \cite{RadhakrishnanCooke1969}.  
Empirical models (e.g. \cite{Rankin1993}) characterize pulsar radio emission 
as having a core beam centered on the magnetic axis and one or more hollow cone beams outside of the core. 
The average-pulse profile widths and component separations are measured to be decreasing functions of radio 
frequency (\cite{Sieber1975}).
This is consistent with a hollow cone beam centered on the magnetic pole, 
emitted at an altitude that decreases
with increasing frequency (radius-to-frequency mapping).
From such profile width measurements, and assuming that the edges of the pulse
are near the last open field line, Kijak \& Gil \cite{KijakGil2003} find an emission radius of
\be  \label{eq:rKG}
r_{\rm radio}  \approx .01\,R_{LC} \left({\dot P\over 10^{ - 15}{\rm s\,s^{-1}}}\right)^{0.07} P^{-0.7} \nu _{GHz}^{ - 0.26} 
\ee
where $\dot P$ is the period derivative.  This
result predicts that for pulsars with $P \lsim 0.1$ s the emission radius $r_{\rm radio}  \gsim 0.05 -0.1\,R_{LC}$.
Such a picture is independently supported by polarization studies \cite{JohnstonWeisberg2006} of pulsars 
younger than 75 kyr.  They concluded that the emission of these pulsars was
from a single wide cone beam, that core emission was weak or absent, and that the height of the 
cone emission is between 1\% and 10\% of the light cylinder radius.

\subsection{The global picture}

As discussed in Section \ref{sec:caustic}, emission along the last open field lines at altitudes 
$r_{\rm em} \sim 0.2 - 0.8\,R_{LC}$ will form caustics, producing sharp peaks in the pulse profiles.  
According to Eqn (\ref{eq:rKG}), the radio cone emission altitude for pulsars with periods $P \lsim 30-50$ ms,
depending on their $\dot P$, will fall in the range of caustic formation.  If the radio conal emission is
radiated over an extended range of altitude (i.e. a few tenths of $R_{LC}$) then radio caustic peaks
would appear in phase with high-energy caustic peaks.  Such a model would explain the multiwavelength phase
coherence of the profiles of fast pulsars like the Crab and millisecond pulsars, PSR B1821-24 and PSR B1937+21.
The conal emission of slower pulsars would fall below the altitude range of caustic formation and the radio
conal peak(s) or core emission would lead the high-energy caustic peaks, as seen in most of the multiwavelength
profiles of Figure \ref{fig:8}.  In these cases, a single radio peak leading double $\gamma$-ray peaks is most
likely to be the edge of a cone beam, since it is unlikely that a viewing angle crosses the narrow core beam as well as both caustics. For example, the observer viewing angle illustrated in Figure \ref{fig:6}
would cross the outer edge of pole 2 at a phase of $180^\circ$.  Due to the radius-to-frequency mapping of
the cone emission, this picture would predict that the measured width of the radio beam would be smaller at 
higher frequency since the emission is originating at lower altitude, whereas the core emission width would not
be expected to vary with frequency.

Aside from the geometry of the emission at different wavelengths, that is directly relevant to observational characteristics of pulsars, there is the more fundamental question of how the microphysics of acceleration and pair cascades fit into the studies of the pulsar magnetosphere.  The magnetosphere models \cite{Contopoulos1999, Spitkovsky2006} make the assumption 
of ideal MHD (no parallel electric fields) in order to derive the global structure of the magnetic field and currents that are solutions to the so-called pulsar equation \cite{Michel1973} that describes a spinning neutron star with a dipole field.  A notable feature of these models is the formation of a neutral sheet in
the equatorial plane of the spin axis which may provide a source of the return current that flows back to
the neutron star along the last open field line. 
The derived currents in such models do not match the currents that have been assumed
in either polar cap or outer gap accelerator models.  Yet the microphysical acceleration models are needed to describe the physics of the pair creation and to check the consistency of the assumption of ideal MHD conditions over most of the magnetosphere.  At present, it is not clear whether the global models can be adjusted to match the boundary conditions of the acceleration models \cite{Timokhin2006} or the whether the acceleration models can produce the required cascade multiplicity with compatible boundary conditions.  
In the end, self-consistency between global models and acceleration models may require time-dependence or
spatial-dependence of the current flow.  

\section{Open Questions}

Although much of the theoretical picture presented in this article depends on a set of undisputed fundamentals,
such as, that rotation-powered pulsars are neutron stars with large-scale dipole magnetic fields that
are directly measurable from their period derivatives, to within uncertainties in the equation of state.
But the picture also involves a number of underlying assumptions that are less well confirmed or agreed upon.
The most important of these for particle acceleration are the boundary conditions on the charge flow.
Both polar cap and outer gap acceleration models assume that the boundary between the open and closed field line
region is a perfect conductor.  This implicitly assumes that there is some microphysical screening at this 
boundary by charges in the closed-field region, although this process has not been modeled or simulated.
The boundary condition on the electric field and potential at the neutron star depends on the binding energy
of charge in the surface layers.  Calculations of the cohesive force for a solid lattice in magnetic fields 
up to about $10^{13}$ G indicate that charges will not be bound in the surface at temperatures above $\sim 0.5$ MK,
so that most pulsars have surfaces hot enough to allow SCLF accelerators to operate.  However, a number of
pulsars have magnetic fields above $10^{13}$ G, where surface conditions have not been adequately explored.
It is possible that the neutron star surface may be in a liquid rather than solid state in very high magnetic 
fields (\cite{JonesCeperley1996, HardingLai2006}) or that charges may be bound by much stronger forces.  
In this case, high-field pulsars could have vacuum instead of SCLF accelerators.  
 
Models often also assume that the neutron star is a dipole field in calculating the accelerating electric 
potential and field.  This assumption could be faulty at low altitude if there exist higher multipole fields
near the neutron star surface, and is certainly invalid at very high altitude where there are distortions of the 
field due to retardation \cite{DyksHarding2004} and current flow (\cite{MuslimovHarding2005}).  The 
$E_{\parallel}$ in the high altitude slot gap \cite{MuslimovHarding2004a} is only approximate and needs to be calculated 
using field line distortions.  It is possible that the distorted field lines, having less curvature than a dipole, 
will have the effect of removing the sign reversals present in the approximate solutions.  There are also
expected to be cross field particle motions at high altitude that will modify solutions of the electric field.

The possible connection of pair cascades to radio emission morphology is an intriguing
question that needs further investigation.  For example, are the pairs from the near-surface polar cap and 
high-altitude slot gap cascades responsible for the core and cone beams?   Do pairs produced in the outer gaps
produce any radio emission?  Whether polar caps and slot gaps can exist on the same field lines or in the
same pulsar magnetosphere is also not known.

Although some of these questions require more theoretical work, it is hoped that many answers will
come from observations with future detectors.  The Large Area Telescope (LAT) on the Gamma-ray Large Area Space 
Telescope (GLAST), due to launch in 2007, 
will have unprecedented sensitivity and energy resolution for gamma-rays in the range of 30 MeV 
to 300 GeV.  GLAST is therefore expected to provide major advances in the understanding of high-energy emission 
from rotation-powered pulsars \cite{Thompson2006, Harding2007}, 
including an increase in the number of detected radio-loud and radio-quiet gamma-ray 
pulsars, and millisecond pulsars, giving much better statistics for elucidating population characteristics, 
measurement of the high-energy spectrum and the shape of spectral cutoffs and determining pulse profiles for a 
variety of pulsars of different ages. Further, measurement of phase-resolved spectra and energy dependent pulse 
profiles of the brighter pulsars should allow detailed tests of magnetospheric particle acceleration and radiation 
mechanisms.  Third-generation ground-based Air Cherenkov
detectors, such as H.E.S.S. \cite{Hinton2004} in Namibia, MAGIC \cite{Lorenz2005} 
in the Canary Islands and VERITAS \cite{Holder2006} in the US, have begun operation.  
They are sensitive to $\gamma$-rays in the range 50
GeV to 50 TeV and are putting important constraints on pulsar
spectra and emission mechanisms \cite{Aharonian2007}.  Further into the future are
planned X-ray telescopes, such as Constellation X, XEUS, and X-ray
polarimeters, such as AXP, POGO and ACT.  Polarimeters \cite{Weisskopf2006} in particular will be
extremely important in verifying model predictions for different pulsar emission geometries 
through phase-resolved measurements of position angle and polarization percent.
 
%
%
%
%
%

%
%



\printindex
\end{document}